\documentclass[]{pasj01}
\begin{document}
\Received{2019/10/30}
\Accepted{2019/12/04}
%\Published{yyyy/mm/dd}

\title{Doppler Tomographic Measurement of the Nodal Precession of WASP-33b}

%%% begin:list of authors
% Do NOT capitalize all letters in "textsc".
\author{Noriharu \textsc{Watanabe}\altaffilmark{1,2,3}}%\thanks{Example: Present Address is xxxxxxxxxx}}
\email{noriharu.watanabe@nao.ac.jp}

\author{Norio \textsc{Narita},\altaffilmark{1,2,3,4,5}}
\email{norio.narita@nao.ac.jp}

\author{Marshall C. \textsc{Johnson}\altaffilmark{6}}
\email{mjohnson@lco.global}
\altaffiltext{1}{Department of Astronomical Science, The Graduate University for Advanced Studies, SOKENDAI, 2-21-1 Osawa, Mitaka, Tokyo 181-8588, Japan}
\altaffiltext{2}{National Astronomical Observatory of Japan, NINS, 2-21-1 Osawa, Mitaka, Tokyo 181-8588, Japan}
\altaffiltext{3}{Astrobiology Center, NINS, 2-21-1 Osawa, Mitaka, Tokyo 181-8588, Japan}
\altaffiltext{4}{JST, PRESTO, 2-21-1 Osawa, Mitaka, Tokyo 181-8588, Japan}
\altaffiltext{5}{Instituto de Astrof\'{i}sica de Canarias (IAC), 38205 La Laguna, Tenerife, Spain}
\altaffiltext{6}{Las Cumbres Observatory, 6740 Cortona Dr., Ste. 102, Goleta, CA 93117, USA}
%% end:list of authors

%% `\KeyWords{}' always has to be placed before `\maketitle'.
 %Do NOT move this preamble from here!

\maketitle

\begin{abstract}
%Please read ``IMPORTANT NOTICE'' carefully before preparing a manuscript. 
WASP-33b is a retrograde hot Jupiter with a period of 1.2 days orbiting around a rapidly rotating and pulsating A-type star. A previous study found that the transit chord of WASP-33b had changed slightly from 2008 to 2014 based on Doppler tomographic measurements. They attributed the change to orbital precession caused by the non-zero oblateness of the host star and the misaligned orbit. We aim to confirm and more precisely model the precession behavior using additional Doppler tomographic data of WASP-33b obtained with the High Dispersion Spectrograph on the 8.2m Subaru telescope in 2011, as well as the datasets used in the previous study. Using equations of a long-term orbital precession, we constrain the stellar gravitational quadrupole moment $J_{2}=(9.14\pm 0.51)\times 10^{-5}$ and the angle between the stellar spin axis and the line of sight $i_{\star}=96^{+10}_{-14}$ deg. These values update that the host star is more spherical and viewed more equator than the previous study. We also estimate that the precession period is $\sim$840 years. We also find that the precession amplitude of WASP-33b is $\sim$67 deg and WASP-33b transits in front of the host star for only $\sim$20\% of the whole precession period. 
\end{abstract}

\section{Introduction}
Although the number is still not many ($\sim$20), planets around hot ($T_{\mathrm{eff}} > 7000$ K) stars have been discovered by transit surveys like WASP (Wide Angle Search for Planets, \cite{2007MNRAS.375..951C}) and KELT (Kilodegree Extremely Little Telescope, \cite{2007PASP..119..923P}). Despite the small number, we have learned that they tend to have a wide range of projected spin-orbit obliquities (\cite{2018AJ....155..100J}). Because hot stars are generally rapidly-rotating, they make themselves more oblate, which makes their orbital nodal precessions faster.
For planets in near-polar orbits especially, nodal precession can be detected more easily during observations spanning several years.
So far, nodal precessions of two planets have been reported: Kepler-13Ab (e.g. \cite{2018AJ....155...13H}) and WASP-33b (\cite{2015ApJ...810L..23J}, hereafter J+15), both of which satisfied the above conditions.

\begin{table*}[htbp]%table1
  \tbl{Parameters of WASP-33 from the Previous Literature}{
  \begin{tabular}{lcl}
      \hline
      Parameter & Value & reference\\ 
      \hline \hline
      Planetary Parameter\\
      $\lambda_{2008}$ (deg) & $-110.06^{+0.40}_{-0.47}$ & \citet{2015ApJ...810L..23J}\\
      $\lambda_{2014}$ (deg) & $-112.93^{+0.23}_{-0.21}$ & \citet{2015ApJ...810L..23J}\\
      $b_{2008}$ & $0.218^{+0.011}_{-0.029}$ & \citet{2015ApJ...810L..23J}\\
      $b_{2014}$ & $0.0840^{+0.0020}_{-0.0019}$ & \citet{2015ApJ...810L..23J}\\
      $R_{p}/R_{s}$ & $0.1143\pm 0.0002$ & \citet{2010MNRAS.407..507C}\\
      $a/R_{s}$& $3.69\pm 0.01$ & \citet{2010MNRAS.407..507C}\\
      $P$ (days) & $1.2198675 \pm 0.0000011$ & \citet{2014A&A...561A..48V}\\
      $T_{c}$ (BJD$_{\mathrm{TDB}}$) & $2456878.65739 \pm 0.00015$ & \citet{2018arXiv181102573V}\\
      $v_{\mathrm{FWHM}}$ (km s$^{-1}$) & $16.2 \pm 0.5$ (TLS) & \citet{2010MNRAS.407..507C}\\
       & $19.2 \pm 0.6$ (McD) & \citet{2010MNRAS.407..507C}\\
       & $18.1 \pm 0.3$ (NOT)& \citet{2010MNRAS.407..507C}\\
      \hline
      Stellar Parameter\\
      $V\sin i_{\star}$ (km s$^{-1}$)&$86.63^{+0.37}_{-0.32}$ & \citet{2015ApJ...810L..23J}\\
      $\log g$ (cgs) & $4.3\pm 0.2$ & \citet{2010MNRAS.407..507C}\\
      $T_{\mathrm{eff}}$ (K) & $7430\pm 100$ & \citet{2010MNRAS.407..507C}\\
      Fe/H & $0.10 \pm 0.2$ & \citet{2010MNRAS.407..507C}\\
      \hline
    \end{tabular}}\label{first}
\begin{tabnote}
Note that TLS, McD and NOT are abbreviations of Th\"{u}ringer Landessternwarte Tautenburg, McDonald Observatory and Nordic Optical Telescope. $v_{\mathrm{FWHM}}$ is FWHM of instinct profile assumed as Gaussian line.
\end{tabnote}
\end{table*}

We focus on the change of WASP-33b's orbit in this study. Table \ref{first} summarizes parameters of WASP-33's system from the previous literature. This planet was first validated with Doppler tomography by \citet{2010MNRAS.407..507C}, who found it is a hot Jupiter orbiting in near-polar retrograde way around an A-type ($T_{\mathrm{eff}}$=7430$\pm$100 K) and rapidly-rotating ($V\sin i_{\star}$ = 85.6 km s$^{-1}$) star. Doppler tomography is one of the methods to measure spin-orbit obliquities and impact parameters simultaneously based on the apparent acceleration of a bump, sometimes referred to as a “planetary shadow”, in the stellar line profiles during planetary transits. This planetary shadow appears in the line profile, because a transiting planet hides a part of its stellar surface and removes spectral contributions to the line profile from the occulted part of the photosphere. J+15 found that the transit chord of this planetary orbit had slightly changed in six years due to its nodal precession. They measured its projected obliquity $\lambda=-110.06^{+0.40}_{-0.47}$ deg and its impact parameter $b=0.218^{+0.011}_{-0.029}$ from the spectral data in 2008, and $\lambda=-112.93^{+0.23}_{-0.21}$ deg and $b=0.0860^{+0.0020}_{-0.0019}$ from the ones in 2014. They then calculated rates of change of these orbital parameters, $\mathrm{d}\lambda/\mathrm{d}t=-0.487^{+0.089}_{-0.076}$ deg yr$^{-1}$ and $\mathrm{d}b/\mathrm{d}t=-0.0228^{+0.0050}_{-0.0018}$ yr$^{-1}$. From the results of J+15, \citet{2016MNRAS.455..207I} later measured the angle between the stellar spin axis and the line of sight $i_{\star} = 142^{+10}_{-11}$ deg and the stellar gravitational quadrupole moment $J_{2}$ $=$ $2.1^{+0.23}_{-0.21} \times 10^{-4}$.

In this paper, we report additional Doppler tomographic measurement of WASP-33b. In Section 2, we summarize our data sets and the methods used to calculate the orbital obliquity and impact parameter from the data sets. Next, we show the results of our analysis in Section 3. We examine how WASP-33b's nodal precession behaves from our results in Section 4. Finally, we present a concludion of this paper in Section 5.

\begin{figure*}[htbp]
 \begin{center}
  \includegraphics[width=\textwidth]{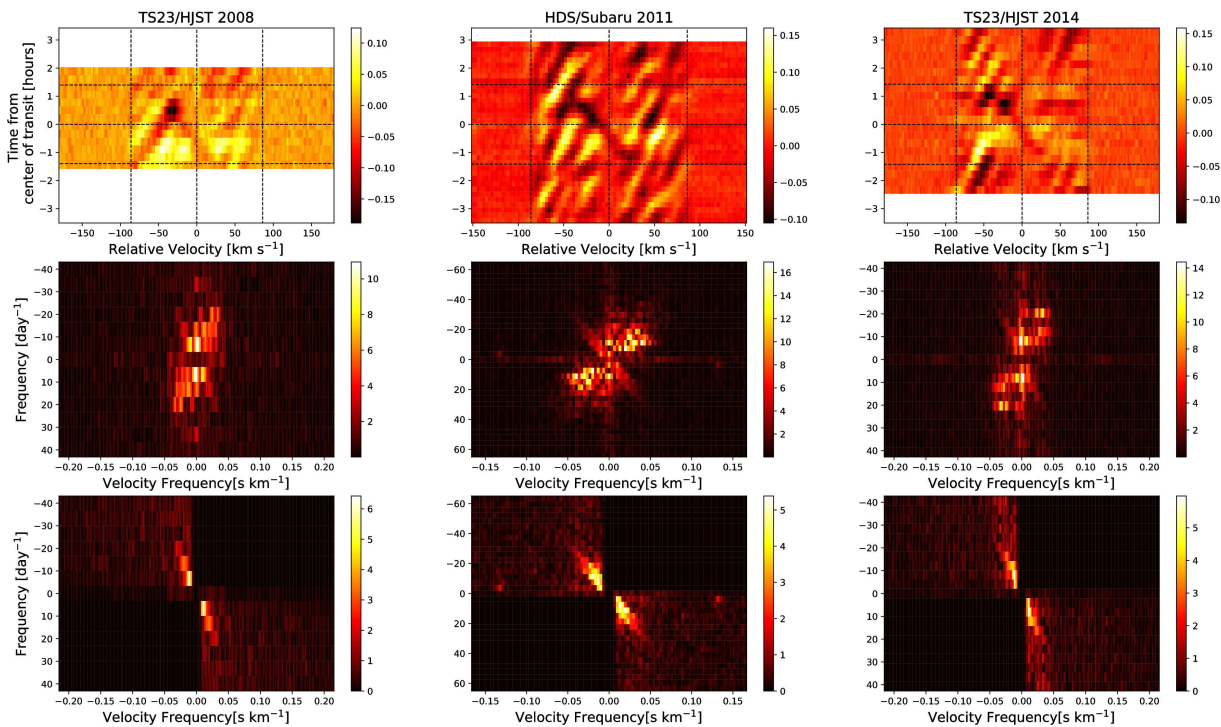}
 \end{center}
\caption{Doppler tomographic data sets and Fourier filters. The first, second and third columns show the data sets of TS23 in 2008, HDS in 2011, and TS23 in 2014, respectively. Top row: observed residuals of line profile series. Virtual dotted lines show $v=0, \pm v \sin i_{\star}$. Bottom, middle and upper horizontal dotted lines show beginning, middle, and end of WASP-33b's transit, respectively. Second row: Fourier spaces after Fourier transform of the residuals of line profile series. These color scales are shown in square-roots. A faint narrow structure from the right bottom to the left upper is a component of WASP-33b's planetary transit. On the other hand, a bright wide structure from the left bottom to the right upper is a component of pulsations. Third row: filtered Fourier space so that only the transit component remains.}
\label{figbff}
\end{figure*}

\section{Methods}
\subsection{Spectroscopic data sets} %2.1
We used three archival spectroscopic data sets of WASP-33 around planetary transits. One of them was taken by the 8.2 m Subaru telescope with High Dispersion Spectrograph (HDS, \cite{10.1093/pasj/54.6.855}) on 2011 October 19th UT. The others are the data sets observed by Harlan J. Smith Telescope (HJST) with Robert G. Tull Coud\'e Spectrograph (TS23; \cite{1995PASP..107..251T}) at McDonald Observatory on 2008 November 12th UT (\cite{2010MNRAS.407..507C}) and 2014 October 4th UT (J+15).

The data set of HDS includes 35 spectra obtained with a resolution of $R=110,000$; 16 spectra taken in-transit. The exposure times are 600 s for 33 spectra and 480 s for 2 spectra. In this study, we adopted a range of wavelength from $4930$\AA\ to $6220$\AA\, except for Na D lines and regions of wavelength around bad pixels. From these spectra, we took continua, corrected them to eliminate the Earth's atmospheric dispersion by dividing spectra of a rapidly rotating star HR8634 ($V \sin I_{\star}\sim 140$~km~s$^{-1}$: \cite{2002ApJ...573..359A}), and shifted these spectra to the barycentric frame. For these processes, we used PyRAF and the calculating tools from \citet{2014PASP..126..838W} and \citet{2010PASP..122..935E}. Then we found that each SNR per pixel of each spectrum was $\sim$160 at $5500$\AA. To pick up each line profile from each spectrum, we adopted least squares deconvolution (LSD; \cite{1997MNRAS.291..658D}). In this method, we regard an observed spectrum as a convolution of a line profile and a series of delta functions. We referred to depths of about 1,000 atomic absorption lines from Vienna Atomic Line Database (VALD; \cite{2000BaltA...9..590K}) and considered these lines as delta functions. Then we derived all of the line profiles and their error bars by the deconvolution using the matrix calculations in \citet{2010A&A...524A...5K}. Finally, we shifted these profiles by the velocity of this system $\gamma=-3.69$ km s$^{-1}$ (\cite{2010MNRAS.407..507C}).

On the other hand, two data sets of TS23 have $R=60,000$ resolution. One data set for 2008 epoch has 13 spectra and SNR per pixel of $\sim$140. The other set for 2014 epoch has 21 spectra and SNR per pixel of $\sim$280. Both of them include 10 in-transit spectra. All of their exposure times are 900 s. We note that the two data sets have been already extracted and published in J+15, and we used the extracted line profile series.

\subsection{Extracting planetary shadow} %2.2
We computed a median line profile for each data set. We subtracted the median line profile from each line profile of each exposure to compute time series of line profile residuals. In the time series of residuals, there are not only a planetary shadow caused by the WASP-33b's transit, but also a striped pattern (see Figure \ref{figbff}). The extra pattern occurs from non-radial pulsations on the surface of WASP-33 (\cite{2010MNRAS.407..507C}). 

To extract only the planetary shadow, we applied a Fourier filtering technique (J+15). First, we did two-dimensional Fourier transform. Second, we made a filter which we set unity in two diagonal quadrants including a power from the planetary shadow, and zero in the other quadrants including a power from the pulsation, with a Hann function between these quadrants. Then, we multiplied the Fourier space by the filter and performed inverse Fourier transform on the filtered Fourier space. These procedures are shown from top to bottom in Figure \ref{figbff}.

\begin{figure*}[htbp]
 \begin{minipage}{0.5\hsize}
  \begin{center}
   \includegraphics[width=80mm]{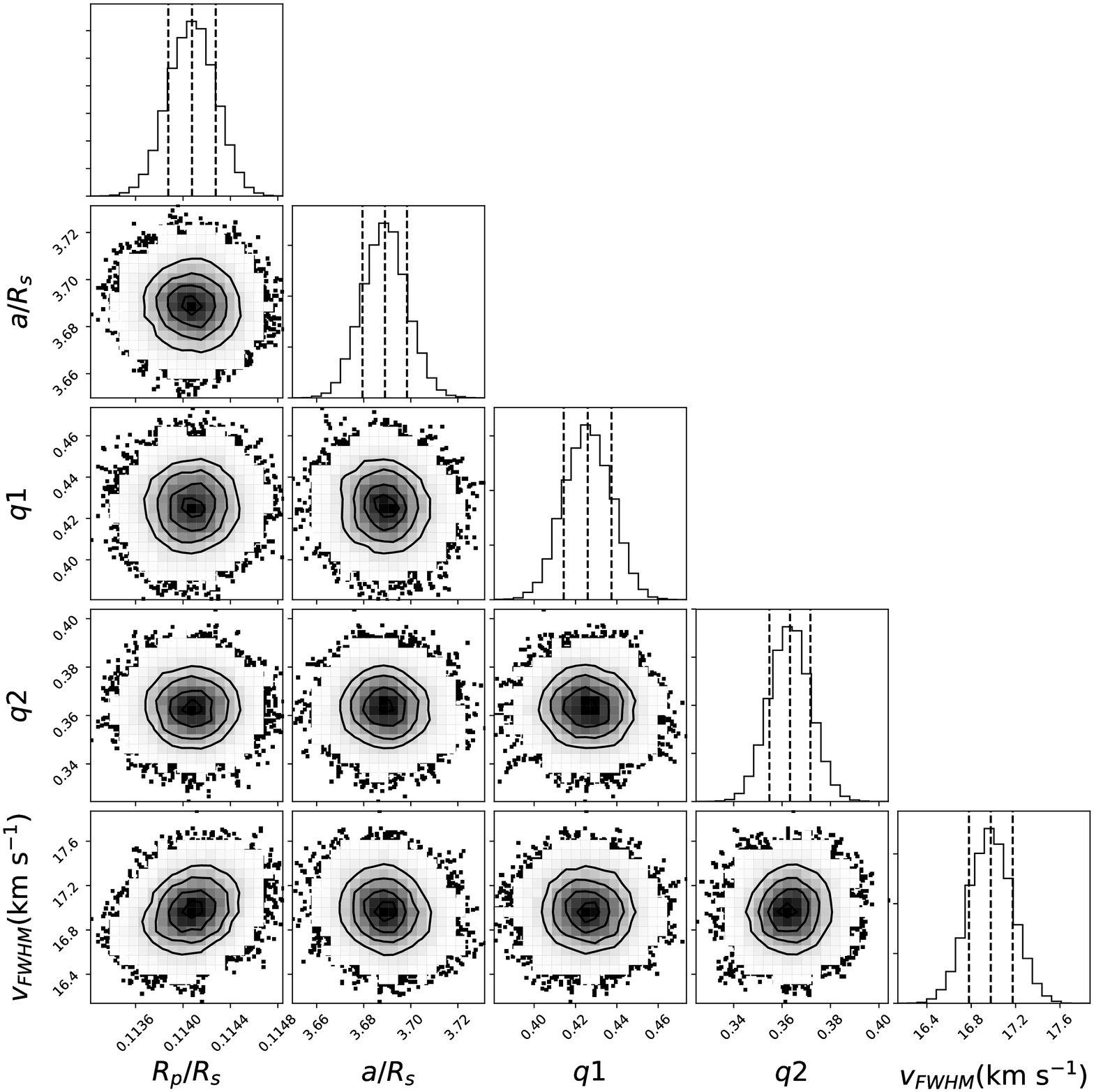}
  \end{center}
 \end{minipage}
 \begin{minipage}{0.5\hsize}
  \begin{center}
   \includegraphics[width=80mm]{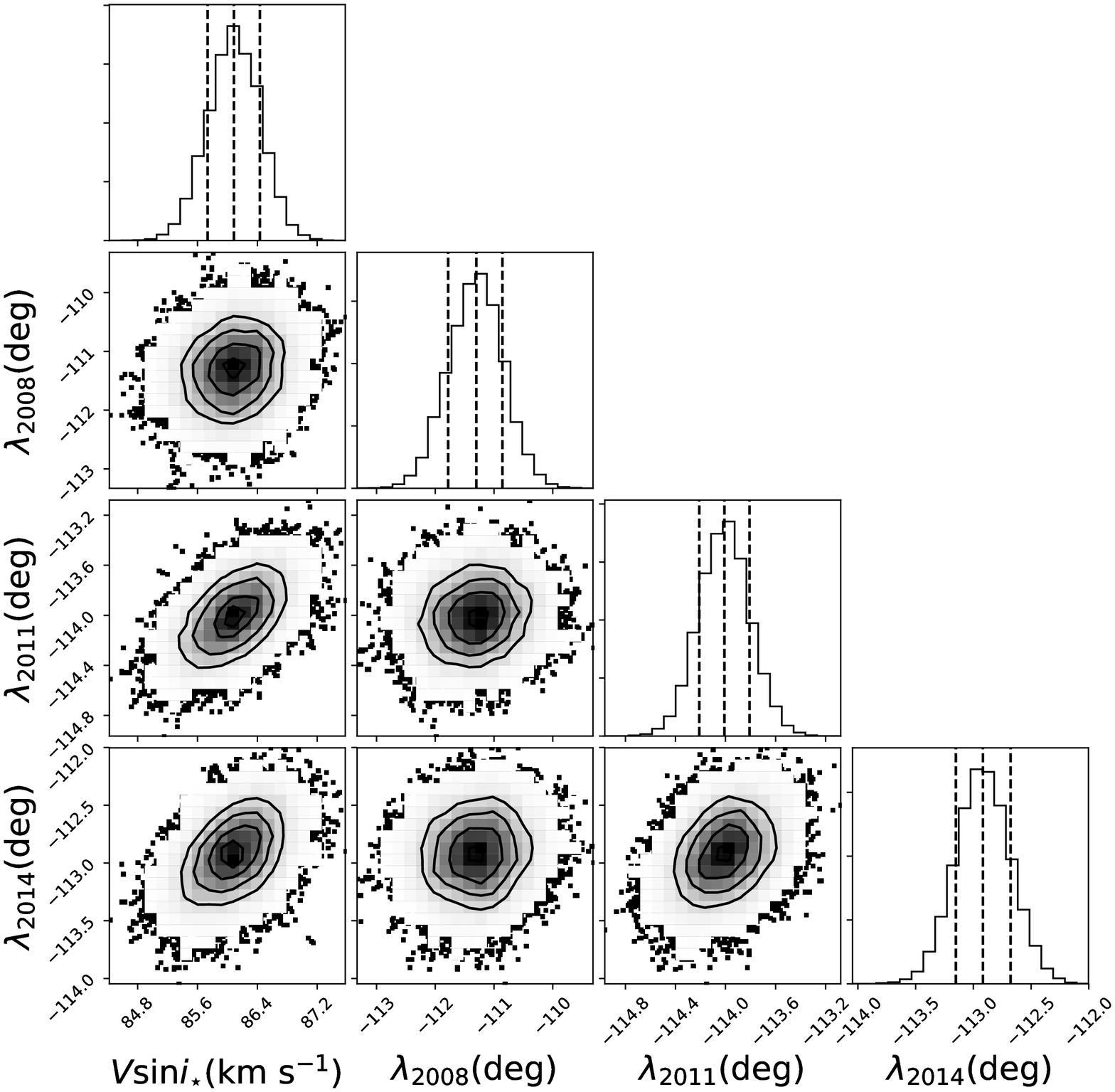}
  \end{center}
 \end{minipage}\\
 \begin{minipage}{0.5\hsize}
  \begin{center}
   \includegraphics[width=80mm]{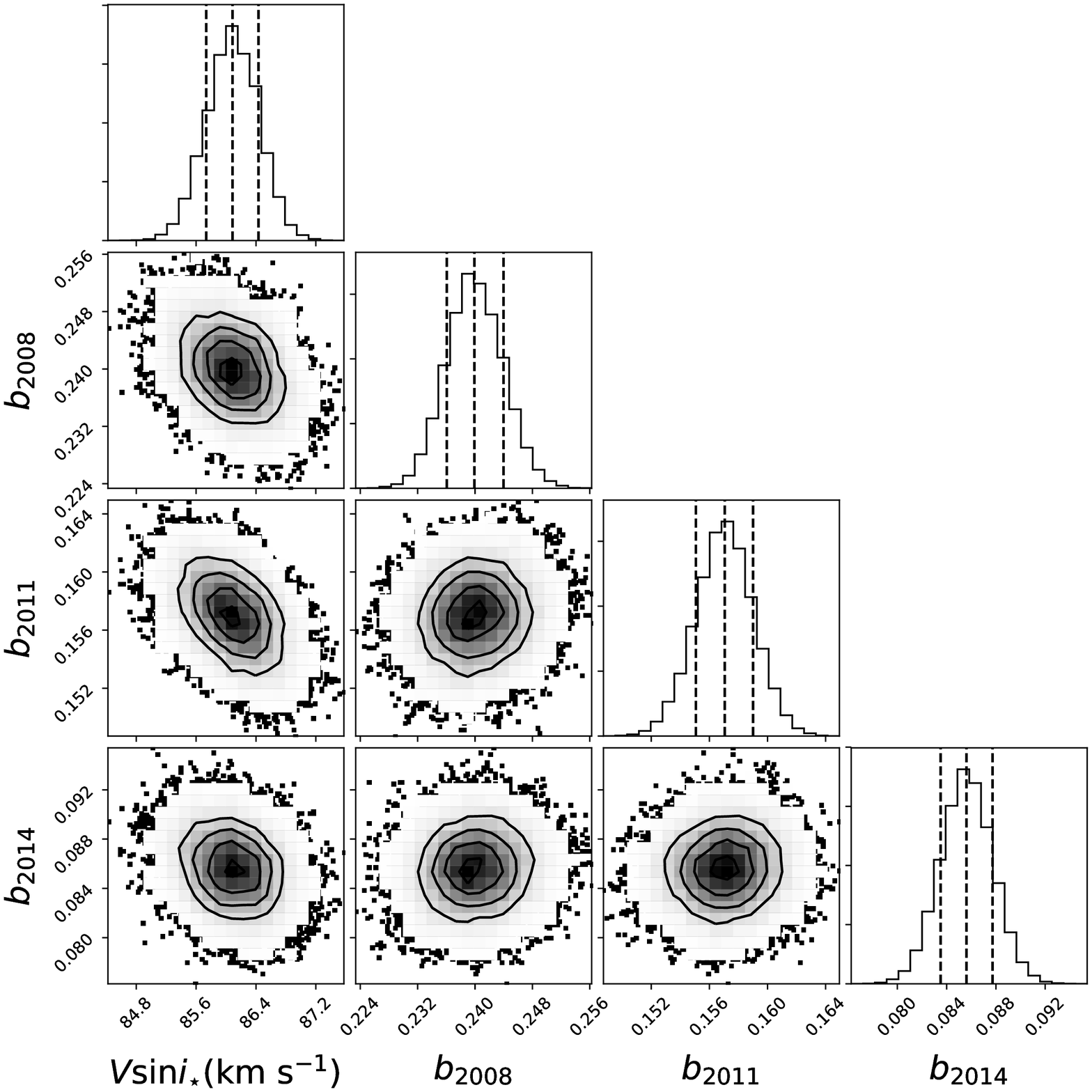}
  \end{center}
 \end{minipage}
 \begin{minipage}{0.5\hsize}
  \begin{center}
   \includegraphics[width=80mm]{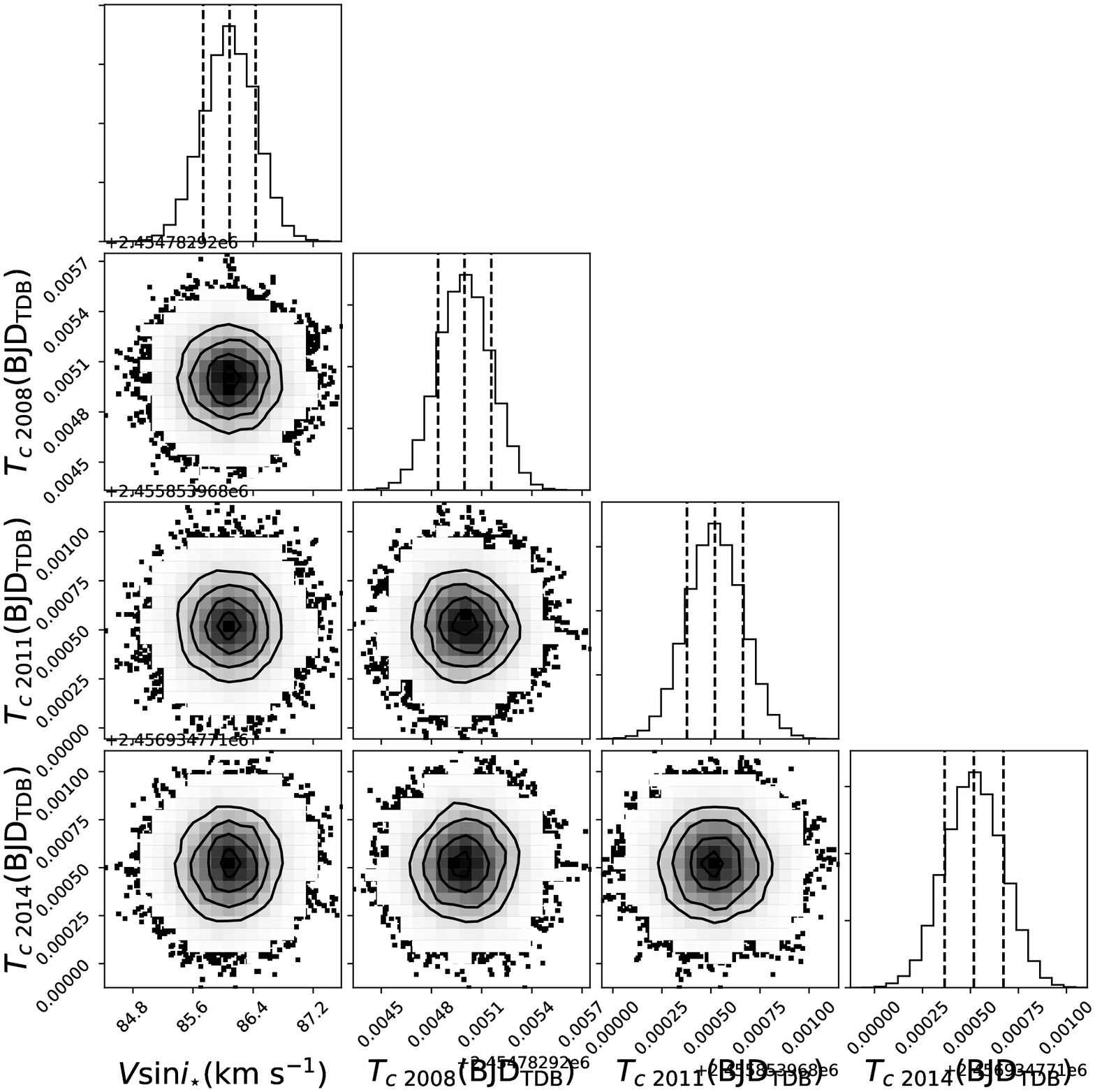}
  \end{center}
 \end{minipage}\\
\caption{Corner plots for the free parameters after using MCMC in section 2.3. Black circles indicate 68\%, 95\% and 99.7\% confidence from the inside. In each posterior distribution of each parameter, vertical dotted lines show its best-fit value (middle) and 1$\sigma$ confidence (both ends). We created these plots with corner.py (\cite{corner}).}
\label{figMCMC}
\end{figure*}

\begin{figure*}[htbp]
 \begin{center}
  \includegraphics[width=\textwidth]{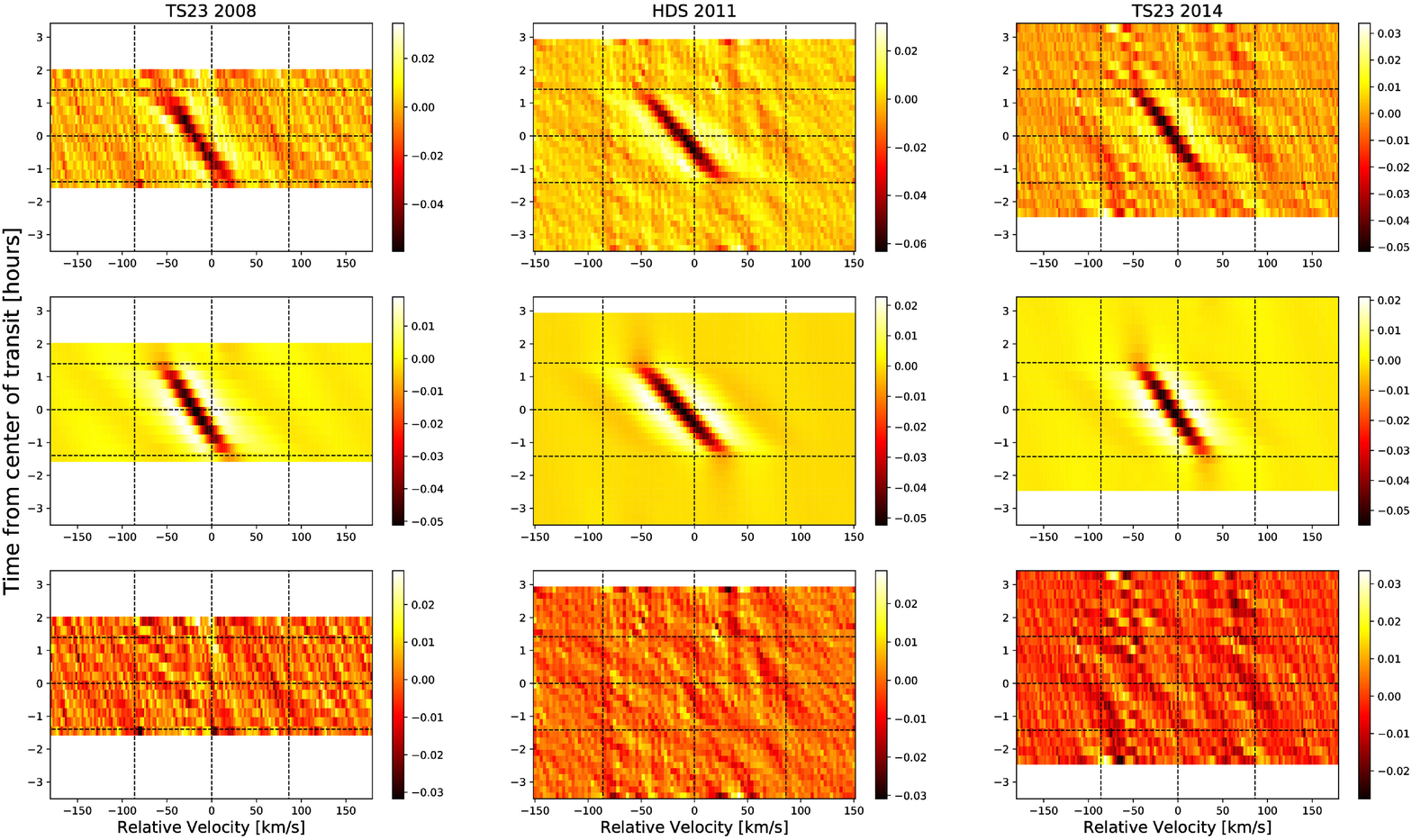}
 \end{center}
\caption{Fitting for filtered residual data by MCMC. These are the same type of colorscale as ones of the first row in Figure \ref{figbff}. First row: residual data remained only a planetary shadow. Second row: filtered models of a planetary shadow using best-fit values. Third row: the difference between the first row and the second row.}
\label{figre}
\end{figure*}

\subsection{Deriving parameters} %2.3

To obtain best-fit values and uncertainties of transit parameters, we adopted Markov chain Monte Carlo (MCMC) using the code $\mathrm{EMCEE}$ (\cite{2013PASP..125..306F}).

We modeled a planetary shadow by convolution between the equations in appendix and a Gaussian line profile due to intrinsic broadening, thermal broadening and micro-turbulence. We then applied the same filter to the planetary shadow model following the procedures described in Section 2.2.

We fitted the observed residuals of the three data sets to the models with 15 parameters using MCMC: $\lambda$, $b$ and $T_{c}$ of each epoch, $V\sin i_{\star}$, $R_{p}/R_{\star}$, $a/R_{\star}$, two quadratic limb darkening coefficients and FWHM of Gaussian line profile. Note that limb darkening coefficients are derived by the triangular sampling method of \citet{2013MNRAS.435.2152K}, $q_{1}$ and $q_{2}$. Here we estimated that $q_{1}$ and $q_{2}$ of HDS and TS23 are equivalent. They can be calculated from the stellar parameters, i.e., effective temperature $T_{\mathrm{eff}}$, surface gravity log $g$, and metallicity. We set priors of $\lambda$ and $b$ for all epochs and the FWHM as uniform functions, otherwise as Gaussian priors. For values and widths of Gaussian priors, we set priors of $R_{p}/R_{\star}$ and $a/R_{\star}$ based on the values and uncertainties from \citet{2013AA...553A..44K}, ones of each $T_{c}$ of each epoch from $P$ in \citet{2014A&A...561A..48V} and $T_{0}$ in \citet{2018arXiv181102573V}, ones of $q_{1}$ and $q_{2}$ calculated by $\mathrm{PyLDTk}$ (\cite{Parviainen2015}, \cite{Husser2013}), and ones of $V\sin i_{\star}$ from J+15.

For the fitting, we minimize the $\chi^{2}$,
\begin{eqnarray}
\label{eq1}
\chi^{2} = \sum_{i} \frac{(O_{i}-C_{i})^{2}}{\sigma^{2}_{i}} + \sum_{j} \frac{(p_{j}-\mu_{j})^{2}}{s^{2}_{j}},
\end{eqnarray}
where $O_{i}$ is the data, $C_{i}$ is the model, $\sigma_{i}$ is the error for the $i$th data point, $p$ is the value of parameter, $\mu$ is the value from the literature, and $s$ is the uncertainty from literature. Indices $j$ and $k$ denote the parameters for the Gaussian priors and the uniform priors, respectively. To converge values of these parameters, We ran 4,000 steps, cut off the first 2,000 steps as burn-in, and iterated this set 100 times. The posterior distributions are plotted in Figure \ref{figMCMC}.

\begin{figure*}[htbp]
 \begin{minipage}{0.5\hsize}
  \begin{center}
   \includegraphics[width=65mm]{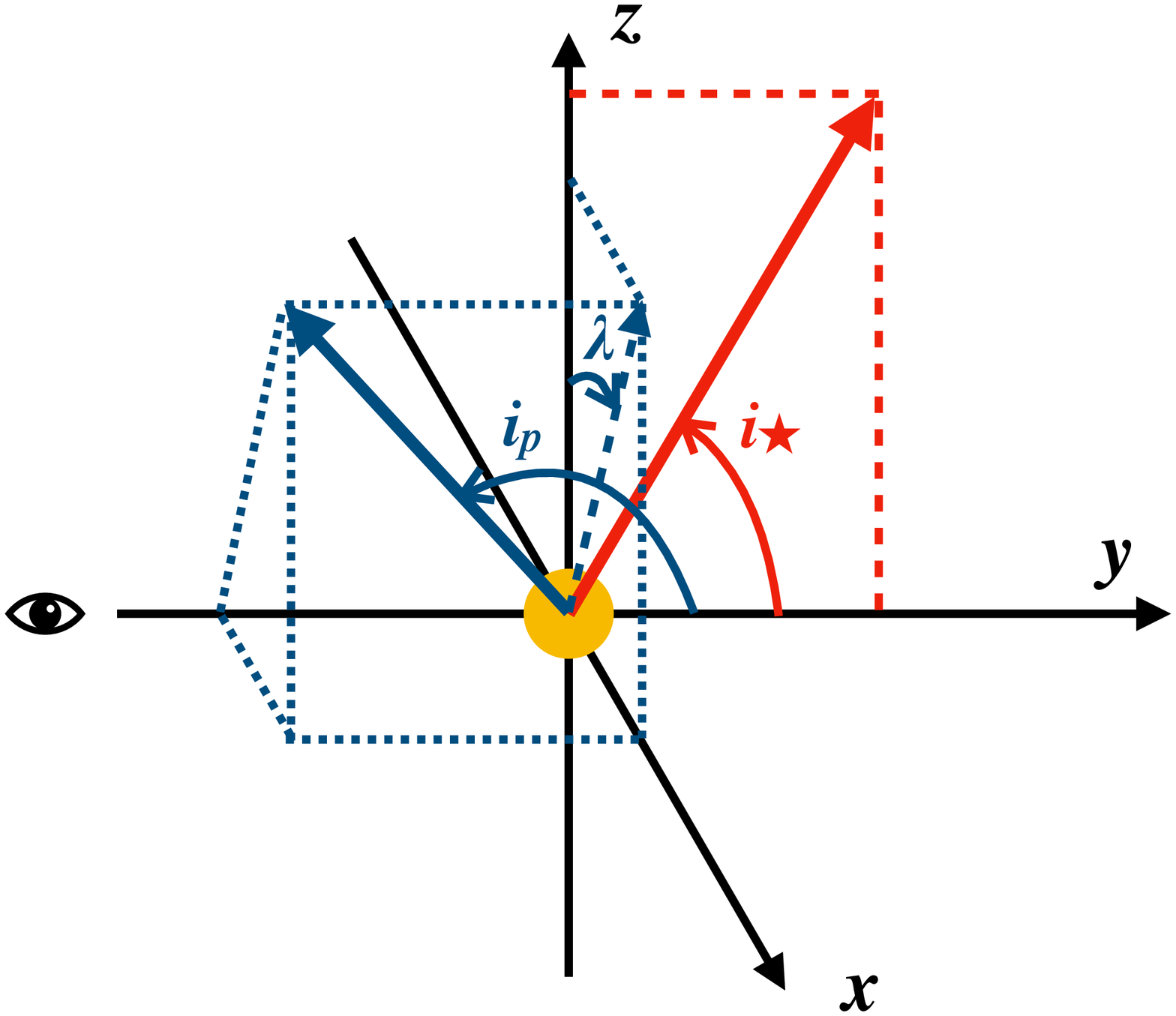}
  \end{center}
 \end{minipage}
 \begin{minipage}{0.5\hsize}
  \begin{center}
   \includegraphics[width=65mm]{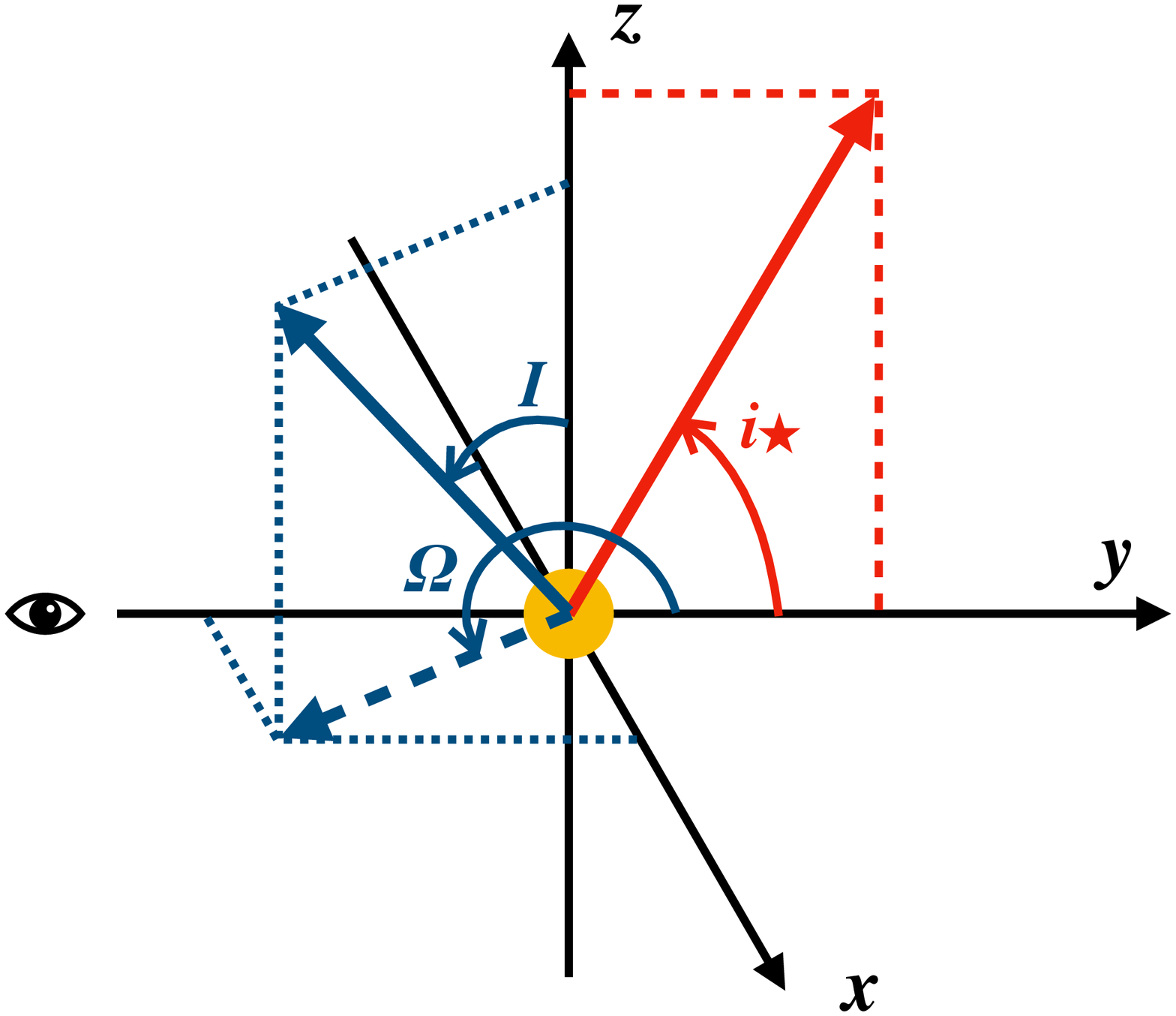}
  \end{center}
 \end{minipage}
 \caption{Outlines of planetary system. We set $y$ axis as line of sight and $xz$ plane as plane of sky. Red vector and blue vector show stellar spin axis and planetary orbital momentum respectively.}
 \label{figpic}
\end{figure*}

\section{Results}
We show the line profile residuals and the best fitted filtered models in Figure \ref{figre}. The best values of $\lambda$ and $b$ are listed in Table \ref{second}. Our results of $\lambda$ and $b$ in 2014 are in excellent agreement with values of J+15, whereas ones in 2008 are marginally consistent with J+15 within $2\sigma$.

We calculated other parameters of angles, $\Omega$, the ascending node, and $I$, the inclination of the orbital plane respect to the $xy$ plane (see Figure \ref{figpic}). These angles can be calculated by using following equations.
\begin{eqnarray}
\label{eq2}
\tan \Omega & = & -\sin \lambda \tan i_{p}, \\
\label{eq3}
\cos I & = & \cos \lambda \sin i_{p},
\end{eqnarray}
where $i_{p}$ is the angle between the line of sight and the momentum of the planetary orbit.
We assumed that $a/R_{\star}$ is constant at any time.
Values of these calculated parameters are shown in Table \ref{second}.

\begin{table}[htbp]
  \tbl{Observed Parameters}{
  \begin{tabular}{lccc}
      \hline
      Parameter & 2008 & 2011 &2014\\
      \hline
      $\lambda$ (deg) & $-111.28^{+0.47}_{-0.48}$ & $-114.01 ^{+0.22}_{-0.20}$ & $-112.91\pm0.24$\\
      $b$ & $0.2397^{+0.0040}_{-0.0039}$ & $0.1571\pm0.0020$ & $0.0856^{+0.0021}_{-0.0020}$\\
      $i_{p}$ (deg)& $86.275^{+0.070}_{-0.072}$ & $87.560\pm0.037$ & $88.671^{+0.034}_{-0.036}$\\
      $\Omega$ (deg)& $86.003^{+0.087}_{-0.091}$ & $87.329\pm0.045$ & $88.557^{+0.039}_{-0.045}$\\
      $I$ (deg)& $111.23^{+0.48}_{-0.47}$ & $113.99^{+0.20}_{-0.22}$ & $112.90 \pm0.24$\\
      \hline
    \end{tabular}}\label{second}
\end{table}

\section{Discussion} %4
The angular momentum of WASP-33b's planetary orbit $L_{p}$ (= $2 \pi M_{p} a_{p}^{2}/P$) is much smaller than the stellar rotational angular momentum of its host star $L_{\star}$; $L_{p}/L_{\star}$ is $\sim 0.05$ using the value of $L_{\star}$ from \citet{2011Ap&SS.331..485I}, ones of $M_{p}$ from \citet{2015AA...578L...4L}, $a_{p}$ and $P$ from \citet{2010MNRAS.407..507C}. In this case, we can regard the stellar rotational axis as a stable vector.
According to Equation (8) and Equation (9) in \citet{2016MNRAS.455..207I}, long-term change of $\Omega$ and $I$ can be expressed as 
\begin{eqnarray}
\label{eq4}
\dot{\Omega} & = & -\frac{3 \pi J_{2} R_{\star}^{2}}{2 a^{2}P}\{2 \sin i_{\star}\cos i_{\star}\cos2I\csc I \cos \Omega \nonumber \\
             & \quad & \quad - \cos I (1-3\sin^{2} i_{\star}+\cos^{2}i_{\star}\cos2\Omega)\},\\
\label{eq5}
\dot{I} & = & -\frac{3 \pi J_{2} R_{\star}^{2}}{a^{2}P}\cos i_{\star} \sin \Omega (\sin i_{\star}\cos I-\cos i_{\star}\sin I\cos \Omega),
\end{eqnarray}
assuming that its orbit is circular and the stellar spin axis stays on $y-z$ plane in Figure \ref{figpic}. These equations need to be solved numerically. Thus we adopt Euler method to fit equations (\ref{eq4}) and (\ref{eq5}) to our results with initial parameters of $\Omega$ and $I$ in 2008 and $\Delta t = 1$ day. In this fitting, we use $\chi^{2}$ fitting method setting $\Omega$ and $I$ at 2008 epoch, $i_{\star}$, and $J_{2}$ as free parameters. We rescale calculated errors of $\Omega$ and $I$ so that $\chi^{2}_{\mathrm{min}}$ equals the number of degree of freedom. Then we estimate errors of $i_{\star}$ and $J_2$ by $\Delta\chi^{2} (\equiv \chi^{2}-\chi^{2}_{\mathrm{min}}) = 1$.

\begin{figure*}[htbp]
 \begin{minipage}{\linewidth}
  \begin{center}
   \includegraphics[width=135mm]{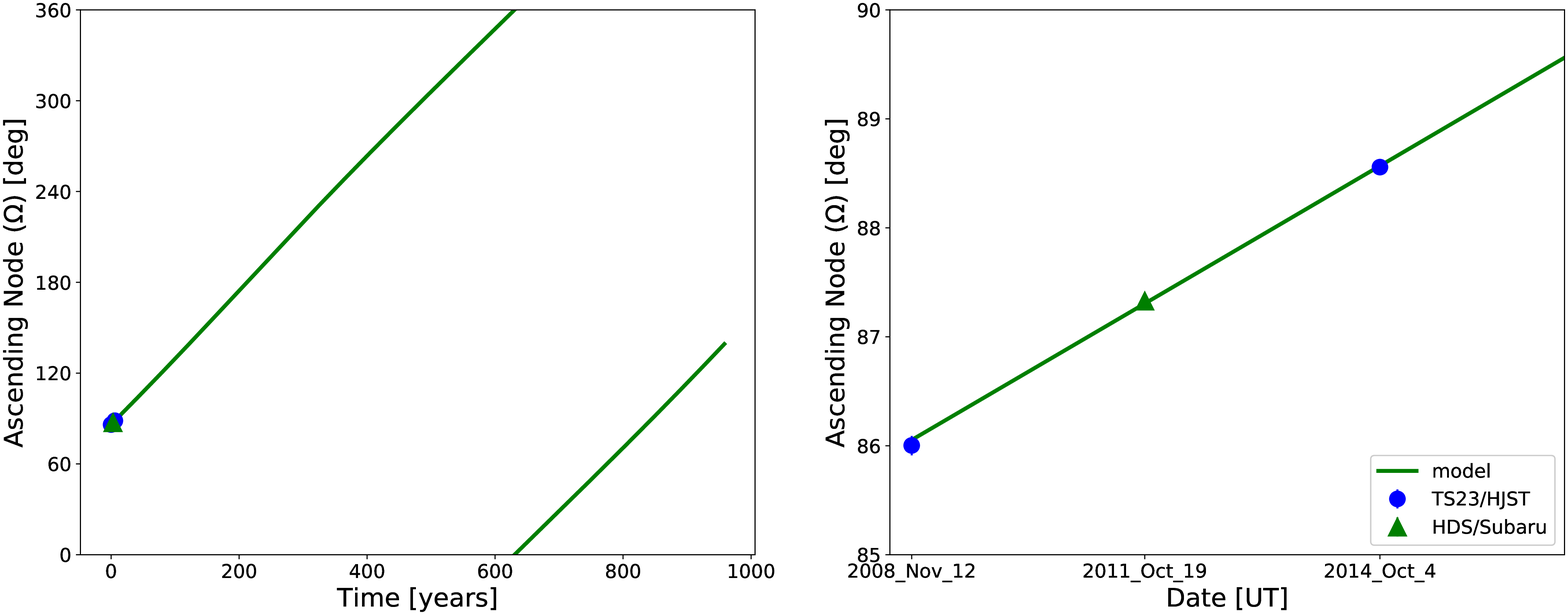}
  \end{center}
 \end{minipage}\\
 \begin{minipage}{\linewidth}
  \begin{center}
   \includegraphics[width=135mm]{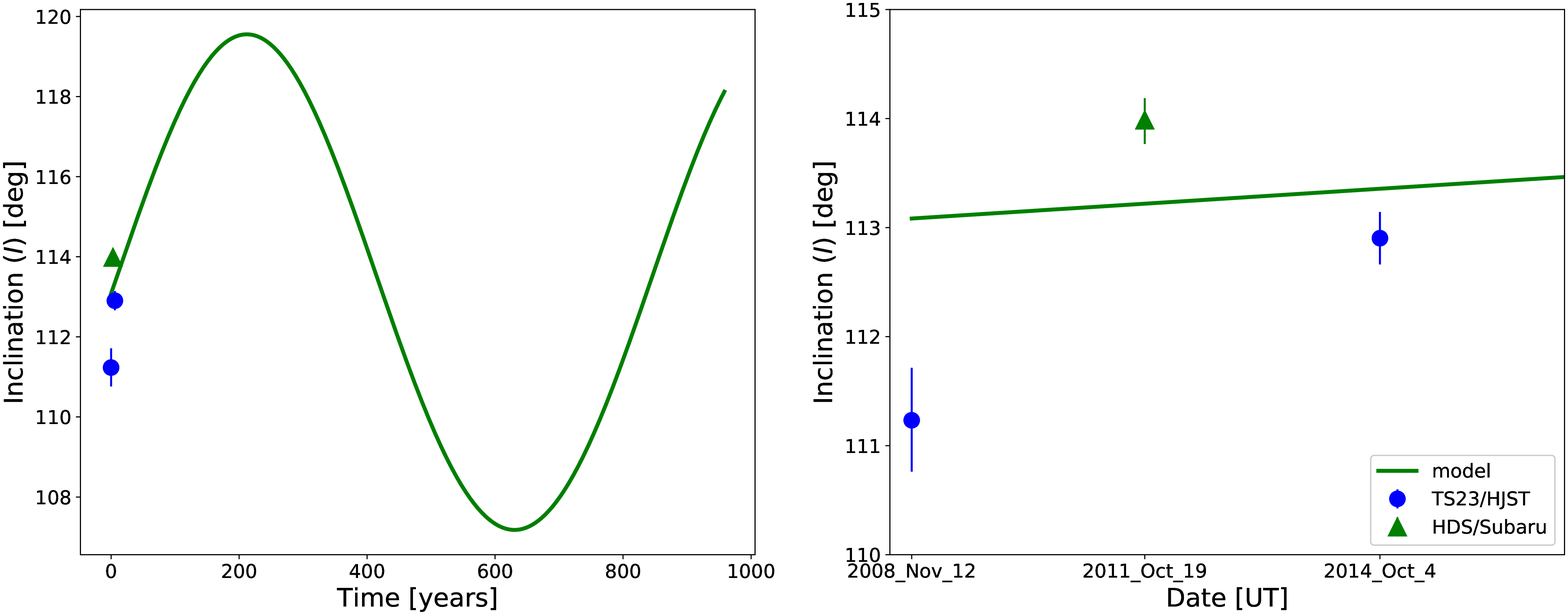}
  \end{center}
 \end{minipage}
 \caption{Changes of $\Omega$ (upper row) and $I$ (lower row). Left and right column shows the change in long term and short term respectively. The x-axis of left column is time in years from the epoch of 2018. The blue circles show values from the data sets of HJST/TS23, whereas the green triangle are values from Subaru/HDS and the green solid lines are model based on \citet{2016MNRAS.455..207I}.}
 \label{OI}
\end{figure*}

Fitted time-variations of $\Omega$ and $I$ in a long-term and a short-term are shown in Figure \ref{OI}. We derive $i_{\star}=96^{+10}_{-14}$ deg and $J_{2}=(9.14\pm 0.51) \times 10^{-5}$, which means WASP-33 is viewed more equator-on and is more spherical than the results in \citet{2016MNRAS.455..207I}. We also estimate that the nodal precession period $P_{\mathrm{op}}$ is $\sim$840 years, which is shorter than one estimated in J+15. Though three data points of $\Omega$ fits the model very well deriving the rate of nodal precession in the short term  $\mathrm{d}\Omega/\mathrm{d}t=0.4269\pm 0.0051$ deg yr$^{-1}$, ones of $I$ does not. This may imply that WASP-33b's precession has a short-term variation or the measured errors of $I$ are underestimated.

\begin{figure*}[htbp]
 \begin{minipage}{\linewidth}
  \begin{center}
   \includegraphics[width=135mm]{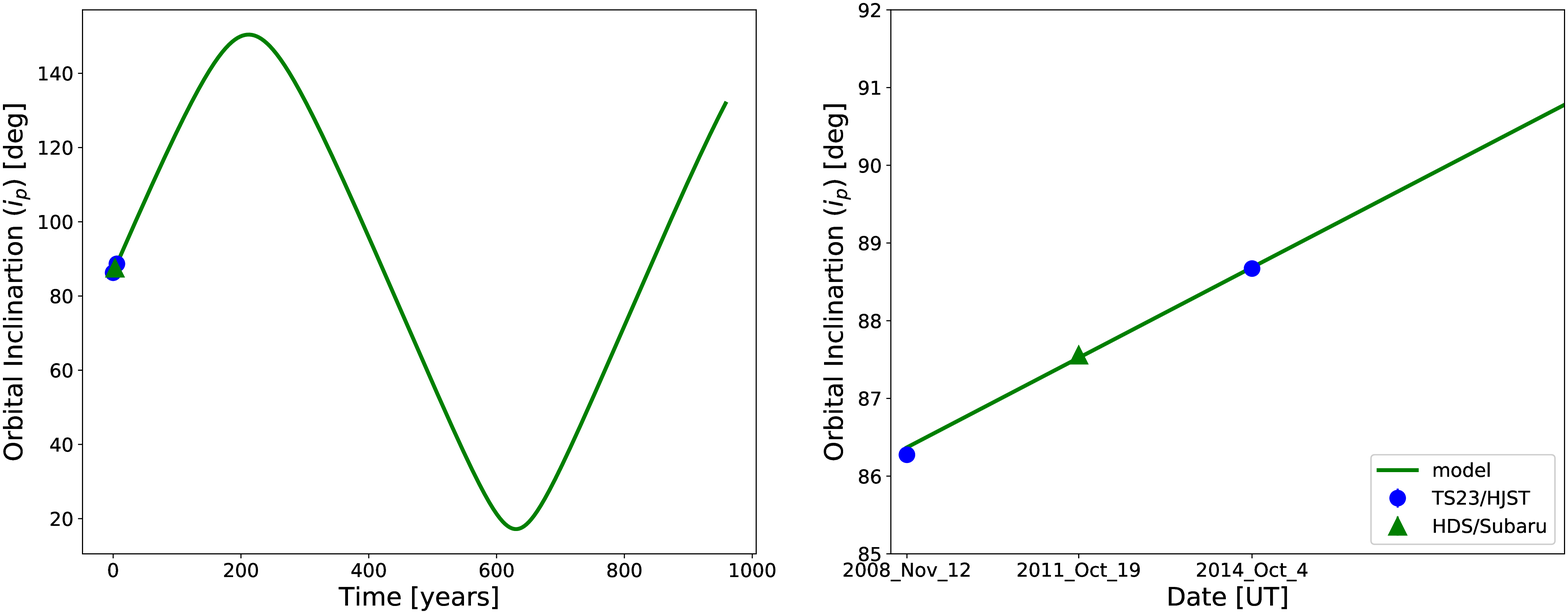}
  \end{center}
 \end{minipage}\\
 \begin{minipage}{\linewidth}
  \begin{center}
   \includegraphics[width=135mm]{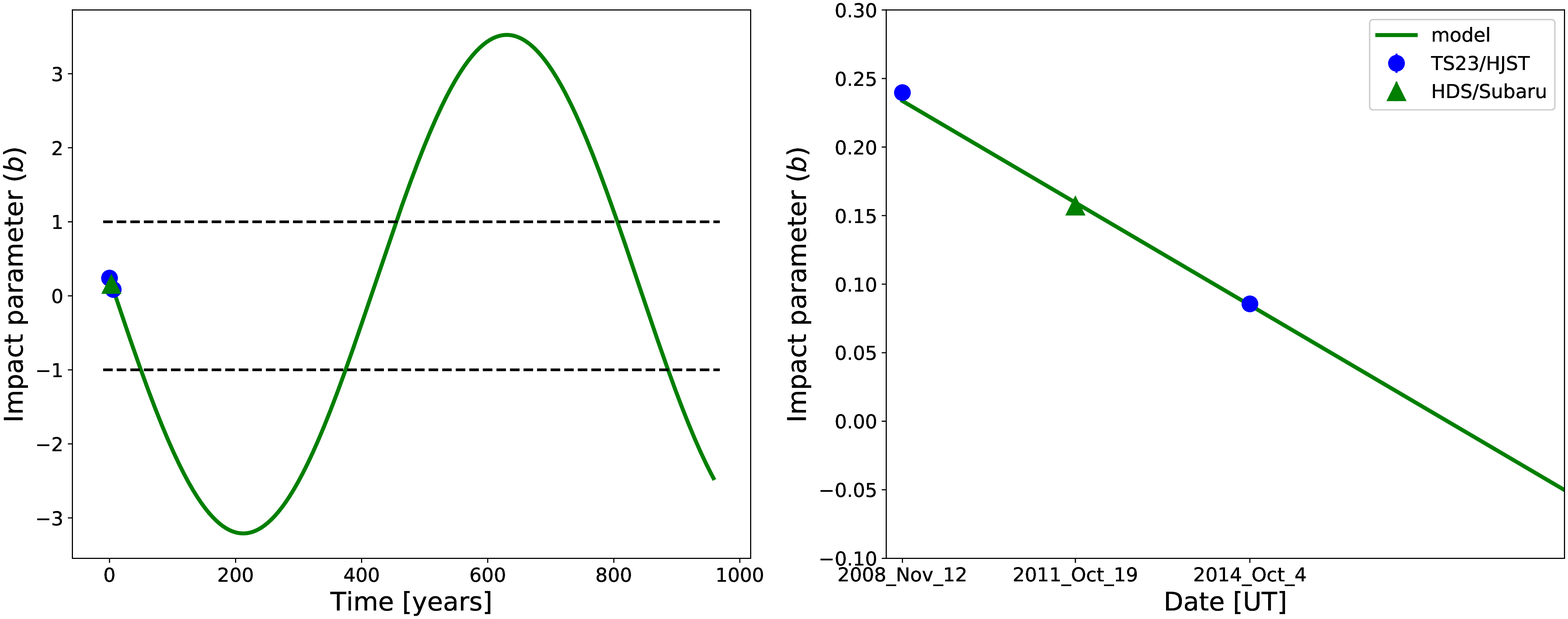}
  \end{center}
 \end{minipage}
 \caption{Changes of $i_{p}$ (upper row) and $b$ (lower row). they are the same types as figure \ref{OI}. In the left bottom figure, two black dotted lines show edges of a stellar disk of WASP-33b. When its impact parameter is between $-1$ and $1$, we are able to see its transit.}
 \label{ib}
\end{figure*}

We also plot changes of orbital inclination $i_{p}$ and impact parameter $b$ in Figure \ref{ib} based on the following equations.
\begin{eqnarray}
\label{eq6}
\cos i_{p} &=& \sin I \cos \Omega \\
\label{eq7}
         b &=& (a/R_{\star}) \cos i_{p}
\end{eqnarray}

We find that the amplitude of WASP-33b's $i_{p}$ is $\sim 67$ deg. We also find that WASP-33b transits in front of the host star for or only $\sim$20\% of the whole precession period, meaning that it is actually rare to discover WASP-33b as a transiting planet. 

\section{Conclusion} %5
We have conducted Doppler tomographic analyses for WASP-33b using archival J+15 datasets from HJST and a new dataset from Subaru in a homogeneous way. We have modeled WASP-33b's precession and derived $i_{\star}=96^{+10}_{-14}$ deg, $J_{2}=(9.14\pm 0.51) \times 10^{-5}$, and the nodal precession timescale to be $\sim 840$ years. We have estimated that the angle between line of sight and WASP-33b's orbital plane oscillates widely with its amplitude of $\sim 67$ deg, and WASP-33b transits its host star only for $\sim20$\% of the period of its nodal precession.

We point out that the nodal precession of hot Jupiters around hot stars would be fairly common, although in varying degrees, since hot stars tend to be oblate due to their rapid rotation and hot Jupiters around hot stars tend to be misaligned. Despite the fact, currently there are only two planets, Kepler-13Ab and WASP-33b, known that show significant nodal precession around A stars. It would also be valuable to measure the nodal precession of another known hot Jupiter, KELT-9b, in the future, because it has a misaligned orbit around a rapidly rotating B/A-type star. \citet{2017Natur.546..514G} indeed guessed that the nodal precession of KELT-9b should be detectable from 2022. In addition, thanks to the ongoing TESS mission, the number of hot Jupiters around B or A-type stars will increase in upcoming years and more planets suitable for this kind of study will become available. As shown in this paper, a measurement of the nodal precession of a hot Jupiter around a hot star can tell us some useful information about its planetary orbit and its host star, such as the fraction of the period when the planet is transiting in front of the host star compared to the whole precession period, or the stellar gravitational quadrupole moment. Future observations of the nodal precession of hot Jupiters around hot stars will reveal the diversity of those points in detail.

\begin{ack}
This paper is based on data collected at Subaru Telescope, which is located atop Maunakea and operated by the National Astronomical Observatory of Japan (NAOJ). We wish to recognize and acknowledge the very significant cultural role and reverence that the summit of Maunakea
has always had within the indigenous Hawaiian community.
The paper includes data taken at The McDonald Observatory of The University of Texas at Austin.

Pyraf is a product of the Space Telescope Science Institute, which is operated by AURA for NASA. This work has made use of the VALD database, operated at Uppsala University, the Institute of Astronomy RAS in Moscow, and the University of Vienna.

This work is partly supported by The Graduate University for Advanced Studies, SOKENDAI, JSPS KAKENHI Grants JP18H01265 and JP18H05439, and JST PRESTO Grant
JPMJPR1775.
\end{ack}

\appendix

\section*{Analytic Model of Planetary Shadow}
We adopt an analytic method of a rotational broadening line profile to create a model of a planetary shadow in Section 2.3. Here we estimate that a host star is a solid sphere and its spin axis is normal to the line of sight though $\sin i_{\star} \neq 1$. This assumption makes $v/V\sin i_{\star}$ equivalent to $x$, a component perpendicular to stellar spin axis in units of the stellar radius (see figure $\ref{figpic}$)

A normalized rotational broadening profile $G(x)$ is given as

\begin{eqnarray}
\label{aeq1}
G(x)=\frac{G'(x)}{\int^{\infty}_{-\infty}G'(x) dx}
\end{eqnarray}
with
\begin{eqnarray}
\label{aeq2}
G'(x)=\int^{z_{\mathrm{max}}}_{-z_{\mathrm{min}}}I(x, z) dz,
\end{eqnarray}
where $I(x, z)$ is a stellar surface intensity at $(x, z)$, the projected position relative to the center of the star. When the intensity is a quadratic limb darkening law, 
\begin{eqnarray}
\label{aeq3}
I(x, z) &=&  I_{0}\left\{1-u_{1}\left(1-\sqrt{1-x^{2}-z^{2}}\right)\right.\nonumber \\
        &\quad& \left. -  u_{2}\left(1-\sqrt{1-x^{2}-z^{2}}\right)^{2}\right\},
\end{eqnarray}
it works out the rotational broadening line profile to the equation
\begin{eqnarray}
\label{aeq4}
G'(x)=\left\{\begin{array}{ll}
\displaystyle 2(1-u_{1}-u_{2})\sqrt{1-x^{2}}\nonumber \\
\displaystyle \quad +\frac{\pi}{2}(u_{1}+2u_{2})(1-x^{2}) \nonumber \\
\displaystyle \quad \quad -\frac{4}{3}u_{2}(1-x^{2})^{3/2} &(|x|<1) \nonumber \\
\nonumber \\
0 &(\mathrm{otherwise})
\end{array} \right. \nonumber \\
\end{eqnarray}
and
\begin{eqnarray}
\label{nor}
\int^{\infty}_{-\infty}G'(x) dx = \pi \left(1-\frac{u_{1}}{3}-\frac{u_{2}}{6}\right).
\end{eqnarray}

When a planet with radius $r(=R_{p}/R_{s})$ is at $(x_{p}, z_{p})$, the projected position written as
\begin{eqnarray}
\label{posix}
x_{p}(t)&=&\frac{a}{R_{s}}\sin\left(2\pi\frac{t-T_{0}}{P}\right)\cos \lambda+b\cos\left(2\pi\frac{t-T_{0}}{P}\right)\sin \lambda \nonumber \\ \\
\label{posiz}
z_{p}(t)&=&-\frac{a}{R_{s}}\sin\left(2\pi\frac{t-T_{0}}{P}\right)\sin \lambda+b\cos\left(2\pi\frac{t-T_{0}}{P}\right)\cos \lambda \nonumber \\
\end{eqnarray}
with time $t$ and covers the stellar disk, the broadening profile is decreased by
\begin{eqnarray}
\label{apla}
K(x, t) &=& \int^{z_{2}(x, t)}_{z_{1}(x, t)}I(x, z) dz \nonumber \\
     & =&  (1-u_{1}-u_{2}(2-x^{2}))(z_{2}-z_{1})+\frac{u_{2}}{3}(z^{3}_{2}-z^{3}_{1}) \nonumber \\
     & \quad& - \frac{u_{1}+2u_{2}}{2}\left[ z_2 \sqrt{1-x^{2}-z^{2}_{2}}-z_1 \sqrt{1-x^{2}-z^{2}_{1}}\right.  \nonumber \\
     &\quad&\quad +\left. (1-x^{2})\left\{\sin^{-1}\left(\frac{z_{2}}{\sqrt{1-x^{2}}}\right)-\sin^{-1}\left(\frac{z_{1}}{\sqrt{1-x^{2}}}\right)\right\}\right] \nonumber \\
\end{eqnarray}
with
\begin{eqnarray}
\label{minus}
z_{1}=\left\{\begin{array}{ll}
0 \quad \quad \quad \quad \quad \left(|x|>1 \ \mathrm{or} \ |x-x_{p}|>r\right) \nonumber \\
\sqrt{1-x^{2}} \quad \quad \left(z_{p}-\sqrt{r^{2}-(x-x_{p})^{2}} > \sqrt{1-x^{2}}\right) \nonumber \\
-\sqrt{1-x^{2}} \quad \left(z_{p}-\sqrt{r^{2}-(x-x_{p})^{2}} < -\sqrt{1-x^{2}}\right) \nonumber \\
z_{p}-\sqrt{r^{2}-(x-x_{p})^{2}}\quad \quad (\mathrm{otherwise})
\end{array} \right. \nonumber \\
\end{eqnarray}
and
\begin{eqnarray}
\label{plus}
z_{2}=\left\{\begin{array}{ll}
0 \quad \quad \quad \quad \quad \left(|x|>1 \ \mathrm{or} \ |x-x_{p}|>r\right) \nonumber \\
\sqrt{1-x^{2}} \quad \quad \left(z_{p}+\sqrt{r^{2}-(x-x_{p})^{2}} > \sqrt{1-x^{2}}\right) \nonumber \\
-\sqrt{1-x^{2}} \quad \left(z_{p}+\sqrt{r^{2}-(x-x_{p})^{2}} < -\sqrt{1-x^{2}}\right) \nonumber \\
z_{p}+\sqrt{r^{2}-(x-x_{p})^{2}}\quad \quad (\mathrm{otherwise})
\end{array} \right. \nonumber \\
\end{eqnarray}
The convolution between $K(x, t)$ and the stellar intrinsic Gaussian profile is our model of the planetary shadow.

Finally, we note that in the appendix of \citet{2015AJ....150..197H}, a part of the above expression
\begin{eqnarray}
z_{2}\sqrt{1-x^{2}-z_{2}^{2}}-z_{1}\sqrt{1-x^{2}-z_{1}^{2}},
\end{eqnarray}
in equation (\ref{apla}) appears to be mistakenly expressed as
\begin{eqnarray}
z_{2}(1-x^{2}-z_{2}^{2})-z_{1}(1-x^{2}-z_{1}^{2}).
\end{eqnarray}

\bibliographystyle{apj}
\bibliography{reference}

\begin{thebibliography}{}
\expandafter\ifx\csname natexlab\endcsname\relax\def\natexlab#1{#1}\fi

\bibitem[{{Abt} {et~al.}(2002){Abt}, {Levato}, \&
  {Grosso}}]{2002ApJ...573..359A}
{Abt}, H.~A., {Levato}, H., \& {Grosso}, M. 2002, \apj, 573, 359

\bibitem[{{Collier Cameron} {et~al.}(2007){Collier Cameron}, {Bouchy},
  {H{\'e}brard}, {Maxted}, {Pollacco}, {Pont}, {Skillen}, {Smalley}, {Street},
  {West}, {Wilson}, {Aigrain}, {Christian}, {Clarkson}, {Enoch}, {Evans},
  {Fitzsimmons}, {Fleenor}, {Gillon}, {Haswell}, {Hebb}, {Hellier}, {Hodgkin},
  {Horne}, {Irwin}, {Kane}, {Keenan}, {Loeillet}, {Lister}, {Mayor}, {Moutou},
  {Norton}, {Osborne}, {Parley}, {Queloz}, {Ryans}, {Triaud}, {Udry}, \&
  {Wheatley}}]{2007MNRAS.375..951C}
{Collier Cameron}, A., {Bouchy}, F., {H{\'e}brard}, G., {et~al.} 2007, \mnras,
  375, 951

\bibitem[{{Collier Cameron} {et~al.}(2010){Collier Cameron}, {Guenther},
  {Smalley}, {McDonald}, {Hebb}, {Andersen}, {Augusteijn}, {Barros}, {Brown},
  {Cochran}, {Endl}, {Fossey}, {Hartmann}, {Maxted}, {Pollacco}, {Skillen},
  {Telting}, {Waldmann}, \& {West}}]{2010MNRAS.407..507C}
{Collier Cameron}, A., {Guenther}, E., {Smalley}, B., {et~al.} 2010, \mnras,
  407, 507

\bibitem[{{Donati} {et~al.}(1997){Donati}, {Semel}, {Carter}, {Rees}, \&
  {Collier Cameron}}]{1997MNRAS.291..658D}
{Donati}, J.-F., {Semel}, M., {Carter}, B.~D., {Rees}, D.~E., \& {Collier
  Cameron}, A. 1997, \mnras, 291, 658

\bibitem[{{Eastman} {et~al.}(2010){Eastman}, {Siverd}, \&
  {Gaudi}}]{2010PASP..122..935E}
{Eastman}, J., {Siverd}, R., \& {Gaudi}, B.~S. 2010, \pasp, 122, 935

\bibitem[{Foreman-Mackey(2016)}]{corner}
Foreman-Mackey, D. 2016, The Journal of Open Source Software, 24,
  doi:10.21105/joss.00024

\bibitem[{{Foreman-Mackey} {et~al.}(2013){Foreman-Mackey}, {Hogg}, {Lang}, \&
  {Goodman}}]{2013PASP..125..306F}
{Foreman-Mackey}, D., {Hogg}, D.~W., {Lang}, D., \& {Goodman}, J. 2013, \pasp,
  125, 306

\bibitem[{{Gaudi} {et~al.}(2017){Gaudi}, {Stassun}, {Collins}, {Beatty},
  {Zhou}, {Latham}, {Bieryla}, {Eastman}, {Siverd}, {Crepp}, {Gonzales},
  {Stevens}, {Buchhave}, {Pepper}, {Johnson}, {Colon}, {Jensen}, {Rodriguez},
  {Bozza}, {Novati}, {D'Ago}, {Dumont}, {Ellis}, {Gaillard}, {Jang-Condell},
  {Kasper}, {Fukui}, {Gregorio}, {Ito}, {Kielkopf}, {Manner}, {Matt}, {Narita},
  {Oberst}, {Reed}, {Scarpetta}, {Stephens}, {Yeigh}, {Zambelli}, {Fulton},
  {Howard}, {James}, {Penny}, {Bayliss}, {Curtis}, {Depoy}, {Esquerdo},
  {Gould}, {Joner}, {Kuhn}, {Labadie-Bartz}, {Lund}, {Marshall}, {McLeod},
  {Pogge}, {Relles}, {Stockdale}, {Tan}, {Trueblood}, \&
  {Trueblood}}]{2017Natur.546..514G}
{Gaudi}, B.~S., {Stassun}, K.~G., {Collins}, K.~A., {et~al.} 2017, \nat, 546,
  514

\bibitem[{{Hartman} {et~al.}(2015){Hartman}, {Bakos}, {Buchhave}, {Torres},
  {Latham}, {Kov{\'a}cs}, {Bhatti}, {Csubry}, {de Val-Borro}, {Penev}, {Huang},
  {B{\'e}ky}, {Bieryla}, {Quinn}, {Howard}, {Marcy}, {Johnson}, {Isaacson},
  {Fischer}, {Noyes}, {Falco}, {Esquerdo}, {Knox}, {Hinz}, {L{\'a}z{\'a}r},
  {Papp}, \& {S{\'a}ri}}]{2015AJ....150..197H}
{Hartman}, J.~D., {Bakos}, G.~{\'A}., {Buchhave}, L.~A., {et~al.} 2015, \aj,
  150, 197

\bibitem[{{Herman} {et~al.}(2018){Herman}, {de Mooij}, {Huang}, \&
  {Jayawardhana}}]{2018AJ....155...13H}
{Herman}, M.~K., {de Mooij}, E.~J.~W., {Huang}, C.~X., \& {Jayawardhana}, R.
  2018, \aj, 155, 13

\bibitem[{Husser {et~al.}(2013)Husser, {Wende-von Berg}, Dreizler, Homeier,
  Reiners, Barman, \& Hauschildt}]{Husser2013}
Husser, T.-O., {Wende-von Berg}, S., Dreizler, S., {et~al.} 2013, A{\&}A, 553,
  A6

\bibitem[{{Iorio}(2011)}]{2011Ap&SS.331..485I}
{Iorio}, L. 2011, \apss, 331, 485

\bibitem[{{Iorio}(2016)}]{2016MNRAS.455..207I}
{Iorio}, L. 2016, \mnras, 455, 207

\bibitem[{{Johnson} {et~al.}(2015){Johnson}, {Cochran}, {Collier Cameron}, \&
  {Bayliss}}]{2015ApJ...810L..23J}
{Johnson}, M.~C., {Cochran}, W.~D., {Collier Cameron}, A., \& {Bayliss}, D.
  2015, \apjl, 810, L23

\bibitem[{{Johnson} {et~al.}(2018){Johnson}, {Rodriguez}, {Zhou}, {Gonzales},
  {Cargile}, {Crepp}, {Penev}, {Stassun}, {Gaudi}, {Col{\'o}n}, {Stevens},
  {Strassmeier}, {Ilyin}, {Collins}, {Kielkopf}, {Oberst}, {Maritch}, {Reed},
  {Gregorio}, {Bozza}, {Calchi Novati}, {D{\rsquo}Ago}, {Scarpetta},
  {Zambelli}, {Latham}, {Bieryla}, {Cochran}, {Endl}, {Tayar}, {Serenelli},
  {Silva Aguirre}, {Clarke}, {Martinez}, {Spencer}, {Trump}, {Joner}, {Bugg},
  {Hintz}, {Stephens}, {Arredondo}, {Benzaid}, {Yazdi}, {McLeod}, {Jensen},
  {Hancock}, {Sorber}, {Kasper}, {Jang-Condell}, {Beatty}, {Carroll},
  {Eastman}, {James}, {Kuhn}, {Labadie-Bartz}, {Lund}, {Mallonn}, {Pepper},
  {Siverd}, {Yao}, {Cohen}, {Curtis}, {DePoy}, {Fulton}, {Penny}, {Relles},
  {Stockdale}, {Tan}, \& {Villanueva}}]{2018AJ....155..100J}
{Johnson}, M.~C., {Rodriguez}, J.~E., {Zhou}, G., {et~al.} 2018, \aj, 155, 100

\bibitem[{{Kipping}(2013)}]{2013MNRAS.435.2152K}
{Kipping}, D.~M. 2013, \mnras, 435, 2152

\bibitem[{{Kochukhov} {et~al.}(2010){Kochukhov}, {Makaganiuk}, \&
  {Piskunov}}]{2010A&A...524A...5K}
{Kochukhov}, O., {Makaganiuk}, V., \& {Piskunov}, N. 2010, \aap, 524, A5

\bibitem[{{Kov{\'a}cs} {et~al.}(2013){Kov{\'a}cs}, {Kov{\'a}cs}, {Hartman},
  {Bakos}, {Bieryla}, {Latham}, {Noyes}, {Reg{\'a}ly}, \&
  {Esquerdo}}]{2013AA...553A..44K}
{Kov{\'a}cs}, G., {Kov{\'a}cs}, T., {Hartman}, J.~D., {et~al.} 2013, \aap, 553,
  A44

\bibitem[{{Kupka} {et~al.}(2000){Kupka}, {Ryabchikova}, {Piskunov}, {Stempels},
  \& {Weiss}}]{2000BaltA...9..590K}
{Kupka}, F.~G., {Ryabchikova}, T.~A., {Piskunov}, N.~E., {Stempels}, H.~C., \&
  {Weiss}, W.~W. 2000, Baltic Astronomy, 9, 590

\bibitem[{{Lehmann} {et~al.}(2015){Lehmann}, {Guenther}, {Sebastian},
  {D{\"o}llinger}, {Hartmann}, \& {Mkrtichian}}]{2015AA...578L...4L}
{Lehmann}, H., {Guenther}, E., {Sebastian}, D., {et~al.} 2015, \aap, 578, L4

\bibitem[{Noguchi {et~al.}(2002)Noguchi, Aoki, Kawanomoto, Ando, Honda,
  Izumiura, Kambe, Okita, Sadakane, Sato, Tajitsu, Takada-Hidai, Tanaka,
  Watanabe, \& Yoshida}]{10.1093/pasj/54.6.855}
Noguchi, K., Aoki, W., Kawanomoto, S., {et~al.} 2002, Publications of the
  Astronomical Society of Japan, 54, 855

\bibitem[{Parviainen \& Aigrain(2015)}]{Parviainen2015}
Parviainen, H., \& Aigrain, S. 2015, MNRAS, 453, 3821

\bibitem[{{Pepper} {et~al.}(2007){Pepper}, {Pogge}, {DePoy}, {Marshall},
  {Stanek}, {Stutz}, {Poindexter}, {Siverd}, {O'Brien}, {Trueblood}, \&
  {Trueblood}}]{2007PASP..119..923P}
{Pepper}, J., {Pogge}, R.~W., {DePoy}, D.~L., {et~al.} 2007, \pasp, 119, 923

\bibitem[{{Tull} {et~al.}(1995){Tull}, {MacQueen}, {Sneden}, \&
  {Lambert}}]{1995PASP..107..251T}
{Tull}, R.~G., {MacQueen}, P.~J., {Sneden}, C., \& {Lambert}, D.~L. 1995,
  \pasp, 107, 251

\bibitem[{{von Essen} {et~al.}(2018){von Essen}, {Mallonn}, {Welbanks},
  {Madhusudhan}, {Pinhas}, {Bouy}, \& {Weis Hansen}}]{2018arXiv181102573V}
{von Essen}, C., {Mallonn}, M., {Welbanks}, L., {et~al.} 2018, arXiv e-prints,
  arXiv:1811.02573

\bibitem[{{von Essen} {et~al.}(2014){von Essen}, {Czesla}, {Wolter}, {Breger},
  {Herrero}, {Mallonn}, {Ribas}, {Strassmeier}, \&
  {Morales}}]{2014A&A...561A..48V}
{von Essen}, C., {Czesla}, S., {Wolter}, U., {et~al.} 2014, \aap, 561, A48

\bibitem[{{Wright} \& {Eastman}(2014)}]{2014PASP..126..838W}
{Wright}, J.~T., \& {Eastman}, J.~D. 2014, \pasp, 126, 838

\end{thebibliography}
%%%

\end{document}